\documentclass[11pt,a4paper]{article}
\usepackage{jheppub}
\usepackage[utf8]{inputenc}
\usepackage{amsmath}
\usepackage{amsfonts}
\usepackage{amssymb}
\usepackage{bbold}
\usepackage{color}
\usepackage{graphicx}
\usepackage{algpseudocode}

\newcommand{\tr}{\mbox{Tr }}
\def\ep{\varepsilon}
\def\half{\frac{1}{2}}
\def\one{\mbox{1 \kern-.59em {\rm l}}}

\title{Eigenvalue-flipping Algorithm for Matrix Monte Carlo
}

\author[1,2]{Samuel Kov\'a\v{c}ik,}
\author[1]{Juraj Tekel}

\affiliation[1]{Department of Theoretical Physics, Faculty of Mathematics, Physics and Informatics, Comenius University in Bratislava, Bratislava, Slovakia}
\affiliation[2]{Department of Theoretical Physics and Astrophysics, Faculty of Science, Masaryk University, Brno, Czech Republic}

\emailAdd{samuel.kovacik@fmph.uniba.sk}
\emailAdd{juraj.tekel@fmph.uniba.sk}

\abstract{Many physical systems can be described in terms of matrix models that we often cannot solve analytically. Fortunately, they can be studied numerically in a straightforward way. Many commonly used algorithms follow the Monte Carlo method, which is efficient for small matrix sizes but cannot guarantee ergodicity when working with large ones. In this paper, we propose an improvement of the algorithm that, for a large class of matrix models, allows to tunnel between various vacua in a proficient way, where sign change of eigenvalues is proposed externally. We test the method on two models: the pure potential matrix model and the scalar field theory on the fuzzy sphere.

\vspace*{.3cm}

\vspace*{.3cm}}

\begin{document}
\maketitle
\section{Introduction}
Matrix models are an important part of both mathematics and physics. The simplest example is the Gaussian Unitary Ensemble that has a wide range of applications \cite{Guhr:1997ve}. Often the action specifying the probability distribution is amended by higher matrix polynomials, multitrace terms, or interaction with fixed matrices motivated by a high energy physics context, see \cite{Szabo:2001kg, Hanada:2013rga, Banks:1996vh, Brezin}.

In some cases, the studied model, or at least its part, can be expressed in terms of matrix eigenvalues. This change of coordinates invokes a term that can be interpreted as logarithmic repulsion between the eigenvalues. Even in the cases where the action is not well suited for this coordinate change, many of the relevant observables depend only on the eigenvalues---it is theretofore useful to think about their behaviour.

Many matrix models have been studied analytically, at least approximately. However, some resist analytic efforts and yield only to numerics. This can be cumbersome as we are often interested in the limit of infinite matrix size and are therefore forced to run simulations with very large matrices \cite{Brezin,Shimamune,Jha:2021exo}.

At least in cases of finite matrix size, models often have a rich structure of vacua between which the system can tunnel. Being able to probe them is important, either for finding the thermodynamically preferred state or as a vital part of the underlying physics. 
 
However, if the vacua are too far apart, the tunneling probability diminishes and it can become difficult for the system to reach the preferred configuration if it has been initiated far away from it. For example, in the studies of the scalar field theory on the fuzzy sphere, which is expressed as a matrix model, it has been observed that the simulated system can become stuck in the false vacuum for sufficiently large matrix size and insufficiently long simulation time. This means that observing transitions between different vacua is notoriously difficult---for example, the uniform order to non-uniform order in the above-mentioned fuzzy field theory case \cite{Ydri:2014rea,Kovacik:2018thy}.

This phase transition is crucial for understanding the consequences of noncommutative structure of the underlying space and the novel features it brings into the field theory defined on such spaces \cite{NCphase}. Asymmetric transitions in matrix models have also been studied in the context of random geometries \cite{Barrett:2015} and the emergence of geometry \cite{Ydri:2021}. Improvement in this regard is thus important for many high-energy physics applications. Different improvements of the traditional Hamiltonian Monte Carlo (HMC) were proposed for the fuzzy sphere model, for example, adding an auxiliary gauge field \cite{Ydri:2014rea}, the overrelaxation method \cite{Panero:2006bx} or alternatively updating either a single matrix element or all of them \cite{GarciaFlores:2009hf}. 
 
In this paper, we propose a simple improvement of the HMC algorithm that overcomes the tunneling issue. The main idea is to include a direct flip of signs of eigenvalues into the algorithm when generating new configurations. This helps the system move between various vacua. This paper is organised as follows. In section \ref{sec2} we discuss the details of the problem with simulating matrix theories with large matrix size and multiple vacua. In section \ref{sec3} we introduce the improved algorithm, which is then tested on a simple pure potential matrix model in section \ref{sec4} and the fuzzy sphere matrix model in section \ref{sec5}. Certain details of eigenvalue transformation and the fuzzy sphere formalism can be found in the appendix.

\section{The tunnelling problem}\label{sec2}
The traditional Monte Carlo method used for matrix models relies on the Metropolis-Hastings algorithm, which produces a Markov chain by proposing new configurations. These new configurations are added to the chain if they are evaluated as improvement, for example by having lower energy or action. They can be accepted even if they are not evaluated as improvement, but only with some probability---the worse they are the smaller it is.

From this stems the trade-off between acceptance rate and decorrelation of configurations in the Markov chain. If the algorithm proposes new configurations that are similar to the previous one, they will be accepted with high probability but are highly correlated.

This issue is partially resolved by using the HMC algorithm \cite{Jha:2021exo,standard,HMC} in which one introduces auxiliary momentum for the system and the configurations are evolved using  Hamiltonian dynamics. This greatly reduces the correlation length. 

However, some issues still persist. This can be easily seen in one of the simplest examples, the random Hermitian matrix model with quartic potential with the probability distribution given by
\begin{align} \label{pure potential}
    \mbox{Prob}(\Phi) = \frac{1}{Z}e^{- N \  \mbox{\small{Tr }} \left( b\Phi^2 + c \Phi^4 \right)}\equiv \frac{1}{Z}e^{-S(\Phi)}\ .
\end{align}
For $-b \gg c >0$, the stable solution is split between two minima at $ \pm \sqrt{-b/2c}$. The goal usually is to compute expectation values
\begin{align}
    \left\langle F(\Phi)\right\rangle=\frac{1}{Z} \int d\Phi e^{-S(\phi)}F(\Phi)\ .\label{expectation}
\end{align}
Following the above mentioned program, this can be done numerically by generating $I$ samples $\Phi_i,\,i=1,2,\ldots,I$ of the ensemble according to probability distribution \eqref{pure potential} and evaluating $\left\langle F(\Phi)\right\rangle=\sum_i F(\Phi_i)/I$. In the large $N$ limit, the above probability distribution is peaked around specific configurations of $\Phi$, which we call vacua, and only configurations not too far from a vacuum contribute to the expectation value \eqref{expectation}. Equivalently, these vacuum configurations are saddle points of the integral, and apart from the saddle with the lowest $S(\Phi)$---with the highest probability---their contribution is suppressed. Due to the $M\to UMU^\dagger$ symmetry of the probability distribution \eqref{pure potential}, the saddles/vacua are properly described in terms of the eigenvalues, see the appendix \ref{appA}.

When simulating this model with finite matrix size $N$, some of the eigenvalues $\lambda_i$ can end up in one potential well while the rest of them in the other
\begin{equation}
\Phi = U \left( \begin{array}{ccc}
\pm \sqrt{\frac{-b}{2c}} & 0 & \ddots  \\ 
0 &\ddots & 0 \\ 
\ddots  & 0 & 
\pm \sqrt{\frac{-b}{2c}}
\end{array}  \right) U^{-1}\ .\label{diagonalization}
\end{equation}

The statistically preferred configuration is in this case the symmetric one with an equal number of eigenvalues in each of the wells. However, if the system thermalizes into an asymmetrical configuration with $\tr \Phi$ oscillating around a non-zero value, it can be difficult for the eigenvalues to tunnel to the other well. 

This introduces another trade-off. Either the momentum impulse in the Hamiltonian procedure is small, the acceptance rate is large and the system covers the vicinity of the given, perhaps false, vacuum. Or the momentum impulse is large enough for the eigenvalues to be able to move to the other well. However, in this case, we need some of the eigenvalues to be updated a lot and some to be updated only a little---so they stay reasonably close to the bottom of their well for the new configuration to be accepted. The difficulty of such well-orchestrated updates grows exponentially with $N$.

In practice, this means that one has to push the system hard during the thermalization phase, reducing the acceptance rate and therefore wasting computational resources, and even then might not be able to make the system tunnel to the true vacuum state.  

\section{Eigenvalue-flipping algorithm}\label{sec3}

To reduce this problem we have established the following algorithm. When proposing a new configuration, before applying the Hamiltonian flow, we first do the eigenvalue decomposition, change the sign of one or more eigenvalues and then compose it back using the same unitary matrix\footnote{The selected eigenvalue is thus teleported into the mirrored position, opposed to tunneling by natural dynamics of the system.}
\begin{equation}
\Phi = U \left( \begin{array}{ccc}
 \lambda_1 & 0 & \ddots  \\ 
0 &\ddots & 0 \\ 
\ddots  & 0 & 
 \lambda_N
\end{array}  \right) U^{-1} \rightarrow \Phi^* = U \left( \begin{array}{ccc}
 \pm \lambda_1 & 0 & \ddots  \\ 
0 &\ddots & 0 \\ 
\ddots  & 0 & 
 \pm \lambda_N
\end{array}  \right) U^{-1}\ .
\end{equation}

The Hamiltonian flow is then applied on $\Phi^*$ which then undergoes the Metropolis-Hastings check against the unflipped configuration before the update. We refer to this algorithm as the eigenvalue-flipping Hamilton Monte Carlo (eHMC). 

Even changing the sign of one of the eigenvalues this way changes many other elements of the original matrix. This change is organized so the overall effect on the matrix action can be small---even zero, for example for an important class of matrix models whose action is an even polynomial in $\Phi$. This has some resemblance to the cluster algorithm for simulating spin systems \cite{hall} as a simple flip of one eigenvalue corresponds to a coordinated change of many matrix elements, as demonstrated in the following example:
\begin{equation*}
\left( \begin{array}{cc}
1 & \frac{1}{2} \\ 
\frac{1}{2} & 0
\end{array} \right) =  U \left( \begin{array}{cc}
\frac{1+\sqrt{2}}{2} & 0 \\ 
0 & \frac{1-\sqrt{2}}{2}
\end{array} \right) U^{-1} \rightarrow  U \left( \begin{array}{cc}
\frac{1+\sqrt{2}}{2} & 0 \\ 
0 & \mathbf{-}\frac{1-\sqrt{2}}{2}
\end{array} \right) U^{-1} = \left( \begin{array}{cc}
\frac{3}{\sqrt{8}} & \frac{1}{\sqrt{8}} \\ 
\frac{1}{\sqrt{8}}  & \frac{1}{\sqrt{8}} 
\end{array} \right) \ .
\end{equation*}

While the computational complexity of the eigenvalue decomposition grows polynomially in $N$, see \cite{pan}, the tunneling becomes exponentially difficult. Of course, some mode allows the eigenvalues to move between the potential wells in small numbers, however, this again becomes a very slow procedure in the large-$N$ limit.  

There are two ways of implementing this algorithm. The first option is to choose an eigenvalue at random and flip it with some probability, we used $p=0.2$. The other is to try to flip each of them individually, for this, we used $p=0.2/N$, so the average number of flipped eigenvalues was the same.

The typical loop in HMC simulations has the following steps: generate new momentum from a heat-bath, store the current configuration, numerically iterate the Hamiltonian evolution, and do the Metropolis-Hastings check. The eigenvalue flipping procedure takes place before the numerical iteration.

There are some scenarios where it makes sense to have the eigenvalue option turned on only during the thermalisation phase, for example, if we know that the system has only a single vacuum in the large-$N$ limit. If we know that for a finite value of $N$ is the same vacuum thermodynamically preferred, the system will thermalize in this state with the highest probability. The eigenvalue flipping procedure can then be turned off at the end of thermalisation phase to reduce the computation effort. 

\section{Comparison for the pure potential matrix model with a quartic potential}\label{sec4}

We will now consider a simple pure potential matrix model with a quartic potential defined in \eqref{pure potential}.

Clearly, as $S(\Phi) = N \sum \limits_{i=1}^N (b\, \lambda_i^2 + c \,\lambda_i^4)$, the action does not depend on all degrees of freedom of the matrix $\Phi$ but only on its eigenvalues $\lambda_i$. However, they are not decoupled as might seem from the action since the path-integration invokes an interaction between them as discussed in the appendix \ref{appA}. 

We will be interested in the case of $b<-2 \sqrt{c}$  where the system is known to have a symmetric two-cut solution in the large-$N$ limit \cite{Shimamune}. Minima of the potential are separated by a distance of $\sqrt{-2b/c}$ and as the parameter $b$ is decreased, their separation increases. 

If one performs HMC simulations of this system starting from a random configuration, the \textcolor{black}{system will termalize into a stable configuration with} eigenvalues split nearly evenly between the two minima. For the purpose of this analysis, to show the benefits of the eigenvalue-flipping algorithm, we instead initialised the system from a configuration in which all eigenvalues reside in the same potential well, that is $\Phi  = \pm \sqrt{-b/2c} \ \mathbb{1}$.

\begin{figure}[t] 
\centering
  \includegraphics[width=1.0\textwidth]{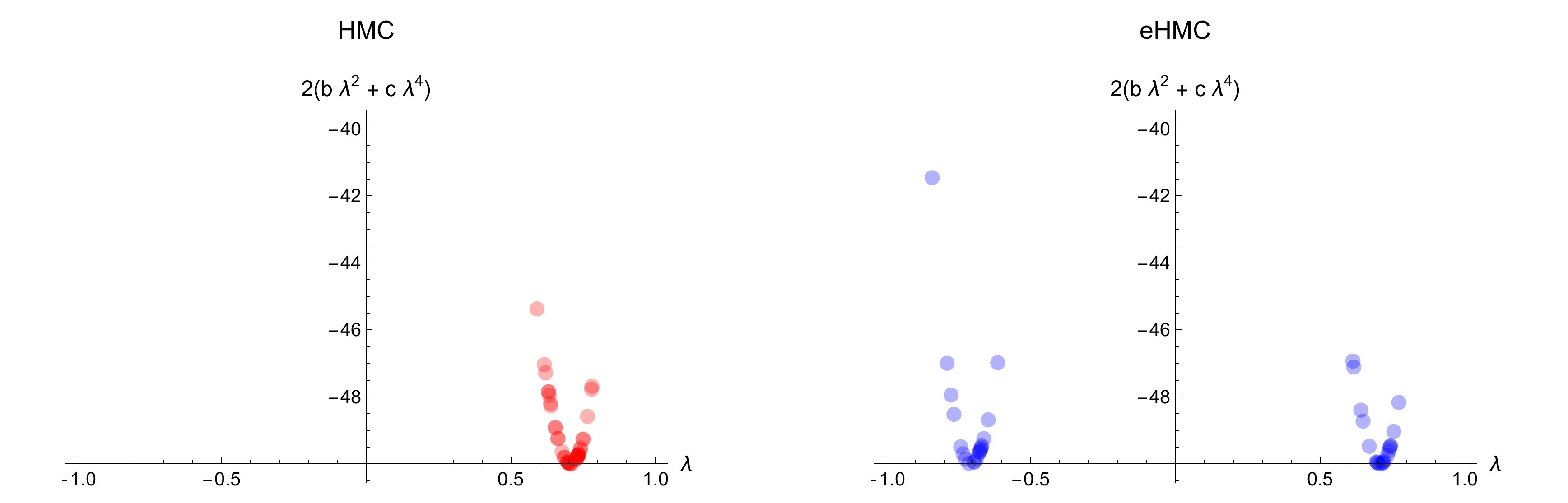}
    \caption{\footnotesize{Eigenvalues of the first 40 configurations for the system with the action \eqref{pure potential} with $N=2$, $b=-100$ and $c=100$. On the vertical axism we show the action corresponding to each eigenvalue. We can see that the eHMC algorithm very quickly probes both potential wells and finds the symmetric solution, despite being initiated in the false (asymmetric) vacuum in which the HMC algorithm becomes stuck.}}
   \label{fig1}
\end{figure}

\begin{figure}[t] 
\centering
  \includegraphics[width=1.0\textwidth]{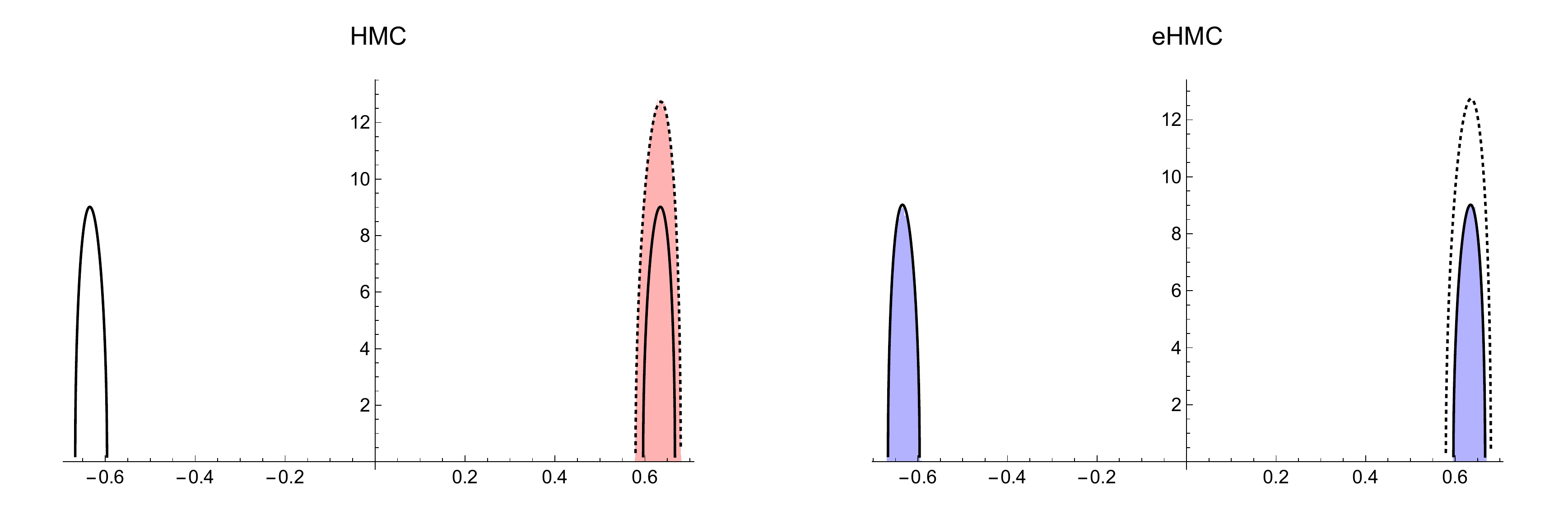}
    \caption{Eigenvalue distributions for the system with the action defined in \eqref{pure potential} with $N=100$, $b=-400$ and $c=100$ produced using approximately $2000$ configuration (HMC) and approximately $1500$ configurations (eHMC) with a similar acceptance rate and simulation running time. The dotted line shows the analytical result for the single-cut asymmetric solution, which is thermodynamically disfavoured (the false vacuum), while the solid line shows the preferred symmetric two-cut solution.}
   \label{fig2}
\end{figure}

Figure \ref{fig1} shows the performance comparison of the ordinary HMC algorithm with the eHMC algorithm after making a small number of steps for a small-$N$ system with separated minima of the potential wells. As we can see in Figure \ref{fig2}, the eHMC algorithm is able to find the preferred vacuum even for rather a large value of ${N=100}$, the solid lines show the eigenvalue distribution obtained analytically in the large-$N$ limit; details can be found, for example, in \cite{Tekel:2015uza}.

Visiting false vacua might be a vital feature of the model. For example, if one is interested in studying a finite-size system, such as in \cite{Pandey:2019dbp}, correct evaluation of the path integral requires probing all vacua. As an example, we can use the matrix model \eqref{pure potential} with $N=10$. There are many relevant vacua, in which some eigenvalues fluctuate around the negative potential well while the rest around the positive one.

\begin{figure}[t] 
\centering
  \includegraphics[width=1.0\textwidth]{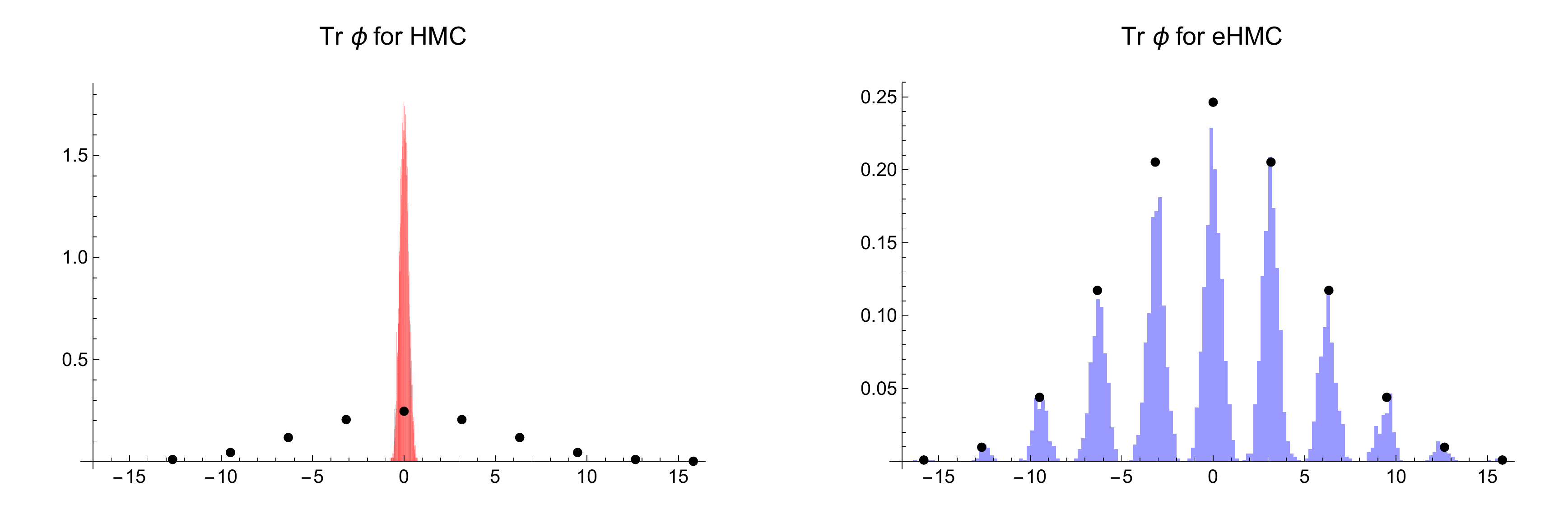}
    \caption{Probability distribution of $\tr \Phi$ obtained from  simulations with the action \eqref{pure potential} with $N=10$, $b=-5$ and $c=1$. It shows $\tr \phi$ for approximately 5000 different configurations. The black points show the binomial probability of having $k$ of $N$ eigenvalues in the same well, their positions have been shifted to mark their corresponding positions. We can see that the eHMC was able to visit all accessible vacua with corresponding probabilities.}
   \label{fig3}
\end{figure}

To see this, we can compute the value of $\tr \Phi$ for each configuration from the Markov chain. In figure \ref{fig3} we can see the comparison for the pure potential matrix model \eqref{pure potential} simulated by both the HMC and the eHMC algorithms. As we can see, the eigenvalue-flipping procedure allowed the system to sample over all important parts of the configuration space. 

\begin{figure}[t] 
\centering
  \includegraphics[width=0.75\textwidth]{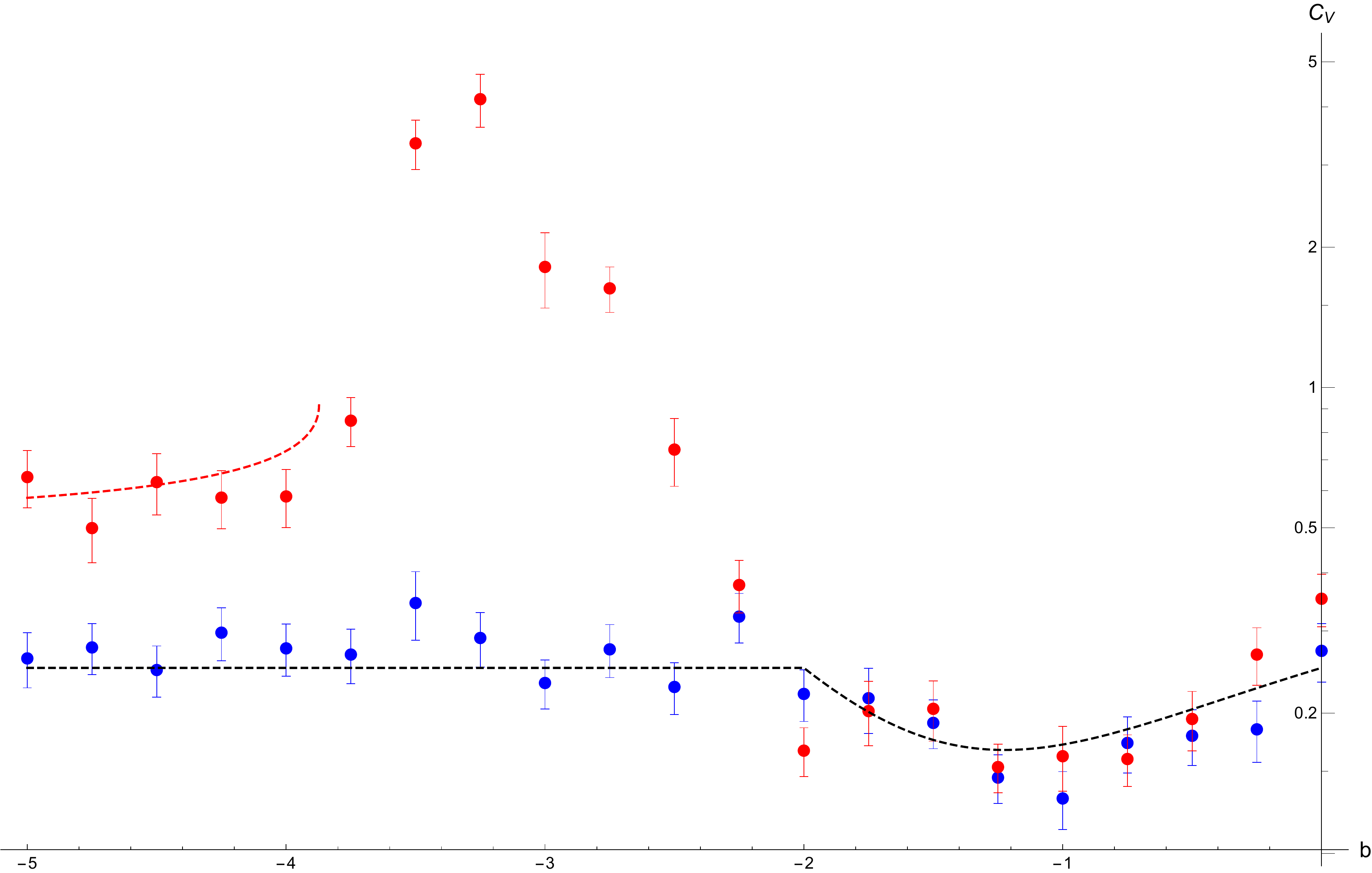}
    \caption{Comparison of the specific heat, $C_V = ( \left\langle S^2 \right\rangle - \left\langle S \right\rangle^2 )/N^2 $, for the model \eqref{pure potential} obtained from very short runs with $10^4$ configurations with $N=40,\, c=1$. The system was initiated from the state $\Phi=\sqrt{-b/2c} \ \mathbb{1}$. The black dashed lines show the analytical result. The red line shows the prediction for the asymmetric phase that is thermodynamically disfavoured, this solution exists only for $b<-\sqrt{15}\ c$.  The black line shows the specific heat for the thermodynamically preferred symmetrical solution. We can see that the eHMC algorithm was able to produce reasonably good results even for very short simulations that were initiated from the false vacuum states. The error bars are the bootstrap estimates.}
   \label{fig4b}
\end{figure}

In Figure \ref{fig4b} we show the comparison of the specific heat computed using very short simulations with both the HMC and the eHMC method. For $b<-2$, the simulations were initiated close to the false vacuum asymmetric state. Compared with the analytical prediction we can observe that the eHMC was able to be reasonably close---within the statistical error---to the analytical large-$N$ prediction. 

\section{Comparison for the scalar field theory on the fuzzy sphere}\label{sec5}

In the previous section, we have shown that the eHMC algorithm was able to tunnel to the preferred vacuum state even with a system with $10^4$ degrees of freedom. A possible suspicion can be that the algorithm is just good at shifting the system towards a symmetric configuration. To disprove this, we will now analyse a model that has a preferred asymmetric solution. A well-known and thoroughly researched example is the fuzzy sphere field theory model, see \cite{Hoppe:1982phd,Madore:1991bw}, with the action given by 
\begin{equation} \label{fuzzy sphere action}
S(\Phi) = N \  \mbox{\small{Tr }} \left( \Phi [L_i,[L_i,\Phi]]  + b\ \Phi^2 + c\ \Phi^4 \right),
\end{equation}
where $L_i, i=1,2,3$ are finite-size representations of the $SU(2)$ generators. Details can be found in the appendix \ref{appB}, from a technical point of view this only presents an additional term in the action. It has been shown before that, in addition to the symmetric one-cut and symmetric two-cut solutions, this model has an asymmetric solution for sufficiently large values of $-b$ \cite{NCphase,Kovacik:2018thy}.

\begin{figure}[t] 
\centering
  \includegraphics[width=1.0\textwidth]{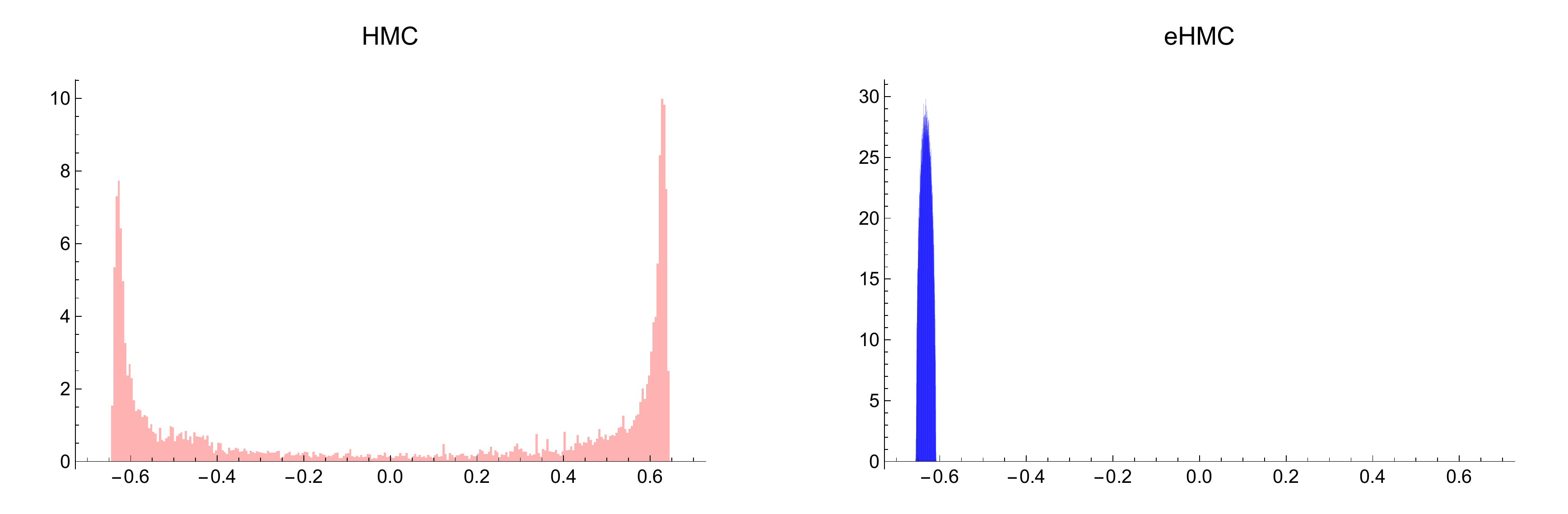}
    \caption{Eigenvalue distribution for first approximately $10^3$ configurations, with thermalisation steps dropped, of the fuzzy sphere model \eqref{fuzzy sphere action} with $N=100$, $b=-400$ and $c=500$. From previous studies \cite{Kovacik:2018thy}, it is known that the asymmetric (with respect to the zero value) configuration is thermodynamically preferred and it has been obtained quickly using eHMC method. The system was initialised from a random configuration in which the eigenvalues are distributed randomly between negative and positive values.}
   \label{fig4}
\end{figure}

The phase diagram of this model has been established in numerous studies \cite{Kovacik:2018thy, Ydri:2014rea, Tekel:2017nzf,MSJT2020}. An important feature is that it has a transition between the symmetric two-cut and asymmetric one-cut solutions (sometimes also called the non-uniform order to uniform order transition). If a simulation of this system is initialized from a random configuration, the eigenvalues will be split evenly between the two wells. The newly introduced term $\mbox{\small{Tr }}  \Phi [L_i,[L_i,\Phi]]$ makes such configurations disfavoured. For $b$ negative enough, the stable solution is an asymmetric one, where the eigenvalues are gathered in the same potential well. 

Comparison of HMC and eHMC simulations of the fuzzy sphere field theory initialized from a random configuration is shown in Figure \ref{fig4}. The eHMC algorithm thermalized in the asymmetric state which is known from the aforementioned studies to be the preferred one for the chosen set of parameters.

\section{Conlucsion}
The eigenvalue flipping algorithm seems to be working well with models that have either symmetric or asymmetric vacua. Even though the eigenvalue decomposition is of some computational difficulty, the procedure allowed the system to thermalize correctly considerably quicker. This allowed us, even on a personal computer, to thermalize the fuzzy sphere model with ${N=400}$, which has ten times more degrees of freedom than the previous large-$N$ study \cite{Kovacik:2018thy}.

It can be difficult to affirm the ergodicity of a simulation but we have shown that at least in some cases the eHMC algorithm quickly visits parts of the configuration space that are inaccessible by the traditional HMC. This, of course, does not guarantee the ergodicity when simulating more complex systems, but still can serve as a step toward it. 

There is an ambiguity regarding the code---one is free to choose how frequently should the eigenvalues be flipped. There are also two options. The first is to choose an eigenvalue at random and flip it with some probability, for example, $p_1=0.2$. The second is to try to flip each of them, where the probability $p_N=p_1/N$ can be chosen so the average number of flipped eigenvalues is the same. We have tested both of those options and they performed comparably well. Another option is to use an adaptive code, where the flipping probability gradually shrinks during the thermalisation and is eventually turned off.  

Why does the algorithm propose new configurations that are easily accepted? In the case of an asymmetric model---even if the symmetry is broken only spontaneously---the explanation is straightforward. The fuzzy sphere term in \eqref{fuzzy sphere action} prefers asymmetric configurations and therefore the algorithm proposes configurations with lower energy/action \footnote{From the point of view of simulating a theory with no time parameter, these terms can be used interchangeably.}. Why does it work in the case of pure potential matrix model \eqref{pure potential} that is unaltered by the change $\lambda_i \rightarrow - \lambda_i$? The reason is the logarithmic repulsion, which is not explicitly present in the action. It can be interpreted as an entropic force, there are more configurations with the same energy/action available when the eigenvalues are separated more and therefore the Hamiltonian flow will have a higher chance of reaching a configuration that will be accepted by the Metropolis check. 

There were various attempts to study the large-$N$ limit of various matrix models which found it increasingly difficult to study properly the transition between the symmetric and the asymmetric regime, see \cite{Lizzi:2012xy,Kovacik:2018thy}. In cases where either the full or parts of the action exhibit the $\lambda_i \rightarrow -\lambda_i$ symmetry, the eigenvalue eHMC might be a powerful enhancement for the simulations.

Implementation of this algorithm depends on the vacuum structure of the action. We have tested two cases where the change $\lambda_i\to-\lambda_i$ worked well, but the procedure can be adjusted to other models---especially for those with more than two minima or minima with unequal depth---rather straightforwardly. 


\subsection*{Acknowledgment}
This research was supported by VEGA 1/0703/20 grant \emph{Quantum structure of spacetime} and the MUNI Award for Science and Humanities funded by the Grant Agency of Masaryk University. We would like to thank Denjoe O'Connor for a valuable discussion. 

\appendix

\section{Eigenvalue transformation}\label{appA}

Following \eqref{diagonalization}, we introduce the angular and eigenvalue degrees of freedom $\Phi=U\Lambda U^\dagger$, which comes with the Jacobian of transformation and change of integration measure \cite{mehta}
\begin{align}
    \Delta^2(\Lambda)=\prod_{i,j;j<i} |\lambda_i-\lambda_j|^2\ \rightarrow\ d\Phi=\left( \prod_{i=1}^N d\lambda_i\right)dU\Delta^2(\Lambda).
\end{align}
The standard approach is to exponentiate this contribution and introduce an action for the eigenvalues only
\begin{align}
    S(\Lambda)= N b \sum_{i=1}^N\lambda_i^2+N c \sum_{i=1}^N \lambda_i^4-2\sum_{i,j;j<i}\log|\lambda_i-\lambda_j|\ .
\end{align}
The explicit factor of $N$ ensures that the potential terms are $\sim N^2$, the same scaling as the Vandermonde contribution, which consists of $N^2$ terms. This guarantees that finite values of parameters $b,c$ produce eigenvalues in a finite range independent of the value of $N$.

\section{The fuzzy sphere}\label{appB}

Let us very briefly introduce the fuzzy sphere and its scalar field theory mentioned in the section \ref{sec5}. For more details, we refer the reader to the original works \cite{Hoppe:1982phd,Madore:1991bw} and the review \cite{steinacker}. The standard sphere can be defined by coordinate functions ${x_a,a=1,2,3}$ in the three-dimensional space which satisfy the constraints
\begin{align}
    x_a x_a=R^2\ ,\ x_a x_b-x_b x_a=0\ ,
\end{align}
where we have explicitly stressed the commutativity of the functions. The fuzzy sphere $S_N^2$ is defined by deforming this commutation relation
\begin{align}
    \hat x_a \hat x_a=R^2\one\ ,\ [\hat x_a, \hat x_b]=i\theta \ep_{abc} \hat  x_c .\label{SFcoms}
\end{align}
More precisely we construct the algebra of functions generated by the above coordinates and define the fuzzy sphere as the object, on which this algebra acts as an algebra of functions. The trained eye recognizes the commutation relations of the $SU(2)$ algebra and thus $\hat x$'s are rescaled generators $L_i$ in the $N$-dimensional representation. (\ref{SFcoms}) generate the algebra of complex matrices and thus real functions on $S_N^2$ are Hermitian ${N\times N}$ matrices $M$.

This enables us to define a very natural version of the Euclidean quantum field theory on $S_N^2$, since the functional integral is simply a finite matrix integral. We thus define the theory by expectation values
\begin{align}
    \left\langle F\right\rangle=\frac{\int {\color{black}d\Phi}\, F({\color{black}\Phi})e^{-S({\color{black}\Phi})}}{\int {\color{black}d\Phi}\,e^{-S({\color{black}\Phi})}}\ ,
\end{align}
with the fuzzy version of the action
\begin{align}
    S(\Phi)=
	\frac{4\pi R^2}{N}\textrm{Tr}\left(
	\half
	\Phi\frac{1}{R^2} [L_i,[L_i,\Phi]]
	+\tilde{b}\ \Phi^2
	+\tilde{c}\ \Phi^4\right) .
\end{align}
This construction simply translates the derivative and integral from the commutative setting into the fuzzy analogues---the commutator with the coordinate function and the trace \cite{Szabo:2001kg,steinacker}. By rescaling the matrices and the coupling constants we can obtain the form \eqref{fuzzy sphere action}. Naively, this action should reproduce the field theory on an ordinary sphere in the large-$N$ limit, but as has been shown in \cite{Kovacik:2018thy} and elsewhere, taking this limit requires more caution.

\bibliographystyle{abbrv}

\end{document}